\documentclass[conference]{IEEEtran}
\IEEEoverridecommandlockouts
\usepackage{cite}
\usepackage{amsmath,amssymb,amsfonts}
\usepackage{algorithmic}
\usepackage{graphicx}
\usepackage{textcomp}
\usepackage{xcolor}
\usepackage{balance}
\usepackage{hyperref}
\usepackage{todonotes}
\usepackage{booktabs}
\usepackage{siunitx}
\usepackage{multirow}
\usepackage{subcaption}
\usepackage{tcolorbox}

\def\BibTeX{{\rm B\kern-.05em{\sc i\kern-.025em b}\kern-.08em
    T\kern-.1667em\lower.7ex\hbox{E}\kern-.125emX}}
\begin{document}

\title{
Semantic Zoom and Mini-Maps for Software Cities
}

\author{\IEEEauthorblockN{Malte Hansen}
\IEEEauthorblockA{\textit{Department of Computer Science} \\
\textit{Kiel University}\\
Kiel, Germany \\
malte.hansen@email.uni-kiel.de}
\and
\IEEEauthorblockN{Jens Bamberg}
\IEEEauthorblockA{\textit{Department of Computer Science} \\
\textit{Kiel University}\\
Kiel, Germany \\
jdbamberg@outlook.de}
\and
\IEEEauthorblockN{Noe Baumann}
\IEEEauthorblockA{\textit{Department of Computer Science} \\
\textit{Kiel University}\\
Kiel, Germany \\
stu237899@mail.uni-kiel.de}
\and
\IEEEauthorblockN{\hspace{60mm}}
\IEEEauthorblockA{\hspace{60mm}}
\and
\IEEEauthorblockN{Wilhelm Hasselbring}
\IEEEauthorblockA{\textit{Department of Computer Science} \\
\textit{Kiel University}\\
Kiel, Germany \\
        hasselbring@email.uni-kiel.de}
\and
\IEEEauthorblockN{\hspace{35mm}}
\IEEEauthorblockA{\hspace{35mm}}
}

\maketitle

\begin{abstract}
Software visualization tools can facilitate program comprehension by providing visual metaphors, or abstractions that reduce the amount of textual data that needs to be processed mentally.
One way they do this is by enabling developers to build an internal representation of the visualized software and its architecture.
However, as the amount of displayed data in the visualization increases, the visualization itself can become more difficult to comprehend.
The ability to display small and large amounts of data in visualizations is called visual scalability.

In this paper, we present two approaches to address the challenge of visual scalability in 3D software cities.
First, we present an approach to semantic zoom, in which the graphical representation of the software landscape changes based on the virtual camera's distance from visual objects.
Second, we augment the visualization with a miniature two-dimensional top-view projection called mini-map.
We demonstrate our approach using an open-source implementation in our software visualization tool ExplorViz.
ExplorViz is web-based and uses the 3D city metaphor, focusing on live trace visualization.

We evaluated our approaches in two separate user studies.
The results indicate that semantic zoom and the mini-map are both useful additions.
User feedback indicates that semantic zoom and mini-maps are especially useful for large software landscapes and collaborative software exploration.
The studies indicate a good usability of our implemented approaches. However, some shortcomings in our implementations have also been discovered, to be addressed in future work. 

Video URL: https://youtu.be/LYtUeWvizjU
\end{abstract}


\section{Introduction}\label{sec:introduction}
Software visualization approaches can be used for various software engineering tasks.
Scientific works in the field of software visualization have explored the visualization of a multitude of data with several visual metaphors ~\cite{merino2018}.
Since a great amount of time is spent reading source code in software development, tools that reduce the necessary time to comprehend source code can speed up software development~\cite{readabilitycode,Bennett_Rajlich_Wilde_2002,janet2016}.
Software visualization can aid the task of program comprehension through visual metaphors, i.e. abstractions that reduce the amount of text which needs to be mentally processed while still allowing developers to build an internal representation of the software and its architecture in their mind ~\cite{4564449}.
Whenever the amount of information that should be visualized becomes too large, the visualization faces similar challenges as the underlying data in terms of comprehensibility.
This issue is known as visual scalability and making a visualization scalable has been acknowledged as one of the largest visualization research problems~\cite{visualresearchproblems,keim2010visualproblems}.

To visualize complex datasets or add additional information to a visualization such as metrics, 3D software visualization techniques are popular~\cite{teyseyre2009,mueller2015}.
The city metaphor, in particular, is popular in 3D software visualization and has been successfully used to convey the structure and behavior of software systems~\cite{knight1999,codecity,wettel2011}.

In this paper, we present two approaches to address the issue of visual scalability in 3D software cities.
The approaches are made concrete through implementations in our open-source software visualization tool ExplorViz.
First, we add features for semantic zoom, whereby the graphical representation of the software landscape changes based on the virtual camera's distance from visual objects.
Additionally, we augment the visualization with a miniature two-dimensional top-view projection called mini-map.

The remainder of this paper is structured as follows.
In the upcoming section, we present related work in the field of semantic zoom and mini-maps.
Section \ref{sec:background} introduces the status quo of ExplorViz before our additions.
In Section \ref{sec:semantic-zoom}, we introduce semantic zoom as an addition to ExplorViz.
The conceptualization and implementation of the mini-map is presented in Section \ref{sec:minimap}.
Both approaches are evaluated in Section \ref{sec:evaluation}.
Section \ref{sec:conclusion} gives a summary of the paper and outlines potential future work.

\section{Related Work}\label{sec:related-work}

Early related work on ExplorViz is the hierarchical software landscape visualization for system comprehension~\cite{VISSOFT2015hierarchical}. Below, we discuss related works on semantic zoom and mini-maps.

\subsection{Semantic Zoom}
Semantic zoom is a widely researched topic for 2D visualizations, especially considering graph visualizations.

Storey et al. present a visualization technique to explore a nested graph view that conveys information on a software's structure~\cite{storey1997}.
They combine a fisheye-view with the ability to pan and zoom in the visualization.
Magnifying a subgraph via zooming may lead to shrinking or removal of other parts of the graph.
In the presented graph structure, certain nodes can contain source code that is hidden by default and only visible at a certain magnification.
Clicking on a node further increases its size so that the presented source code becomes more readable.
The approach of Storey et al., although not coined as semantic zooming, is an early implementation of semantic zoom in software visualization.

Wiens et al. present semantic zooming for ontology graph visualizations~\cite{Wiens2017}.
The underlying ontology data can lead to graph representations with many intersecting edges and a high degree of visual complexity.
They address this issue by introducing different layers with various level of details.
For example, edges may be aggregated and labels can be hidden.
Wiens et al. conducted a study with 12 participants, who were divided into two groups.
Participants of both groups solved tasks with and without the implemented semantic zoom features.
In the end, participants were asked to rate their experience, including the readability, visual clarity, and
information clarity.
In that study, the semantic zoom approach outperformed the version without semantic zoom.
However, the participants asked for additional indicators, tooltips, and reported back a steep learning curve.

Kasperowski et al. present an approach called KIELER, a framework for automatic diagramming of complex systems~\cite{Kasperowski2024}.
A key feature to enable the exploration of large diagrams is a semantic zoom approach, which the authors call smart zoom.
The approach focuses on hierarchically organized diagrams and adapts the level of detail depending on the current zoom level.
A user only needs to zoom and drag in the diagram to make use of this feature.
Elements in the diagram can be hidden or scaled and important labels or icons are appropriately scaled and positioned to keep them visible when zoomed out.
The semantic zoom approach is showcased with a large state chart diagram.

De Luca et al. present an interactive graph visualization with semantic zoom for multi-level trees~\cite{deluca2019multileveltreebasedapproach}.
Different layers are computed to visualize the graph appropriately for various zoom levels.
Inspired by common visualization techniques for maps, they present seven desirable visual properties for their visualization.
These include common techniques like the minimization of edge crossings or that details may not be hidden when increasing the zoom.

Other approaches include CodeCanvas by DeLine and Kael that introduces semantic zoom to the user interface of code editors~\cite{DeLine2010}.
Köth et al. present an approach for semantic zoom for visual UML editors~\cite{Kth2002}.

In the realm of software visualization, Frisch et al. present an approach to apply semantic zooming techniques to UML diagrams~\cite{Frisch2008}.
They introduce different levels of details and allow to focus visual elements.
In addition, they add visual indicators which support the navigation and orientation when zoomed in.

EvoSpaces visualizes complex software systems as 3D software cities~\cite{Alam2007}.
The authors describe that otherwise textured buildings can be made transparent to reveal a visualization of functions inside the buildings.
In addition, buildings are rendered without textures in a night view that focuses on the visualization of traces.
However, to the best of our knowledge, there is no specific research on semantic zooming techniques for 3D software visualization approaches.

\subsection{Mini-Map}
Mini-maps are a common feature of video games to help users navigate.
Zagata and Medyńska-Gulij have analyzed the mini-maps of 100 popular video games to extract their key attributes~\cite{mini-map-design}.
The result are 8 parameters and attributes that enable the classification of mini-map features.
These parameters namely are shape, position, orientation, centering, projection, base layers, proportions, and additional navigational elements.
In the sample of video games, the shape is typically a circle or rectangular, the mini-map usually placed in the bottom left or top right, the orientation is camera based or static, and the centering is to 80\% player-based.
In addition, the projection is almost always orthographic, artificial base layers are used, and a mini-map usually takes up 1.1-4\% of the screen.
Finally, the mini-map may be augmented with additional elements, mostly including arrows.
For our approach to adapt mini-maps for 3D software cities (see Section \ref{sec:minimap}), we build on these results. 

In the realm of software visualization, FlyThruCode by Oberhauser et al. is an approach to visualize software by using a combination of the universe and terrestrial metaphor~\cite{flythrough}.
In the universe metaphor, each class is represented as a planet, and packages are represented by solar systems.
For the terrestrial metaphor, planets can be visited with a spaceship and buildings, representing classes on the planet, may be inspected.
The terrestrial view of FlyThruCode uses a small mini-map in the top right corner.
Due to its small size, it must be clicked to be displayed in an enlarged and readable view.
Our approach is similar, but we have expanded the concept to be more prominent, configurable, and interactive.

Aside from FlyThruCode, we are not aware of similar mini-map implementations in 3D software visualization tools.
Code Park by Khaloo et al. is a visualization approach inspired by video games~\cite{codepark} that offers a bird's view.
This view is similar to the perspective that a mini-map would take but is implemented as an alternative view and not added as part of another view.

CodeMetropolis by Balogh and Beszédes uses the 3D game Minecraft to visualize code~\cite{codemetropolis}.
They mention the plan to implement a mini-map as future work to improve the support for navigation.

A related concept of mini-maps is the polymetric view in software visualization that offers multiple perspectives on the same software~\cite{Lanza2003,anslow2010}.

\section{Background}\label{sec:background}
\begin{figure}[htbp]
	\includegraphics[width=\textwidth/2]{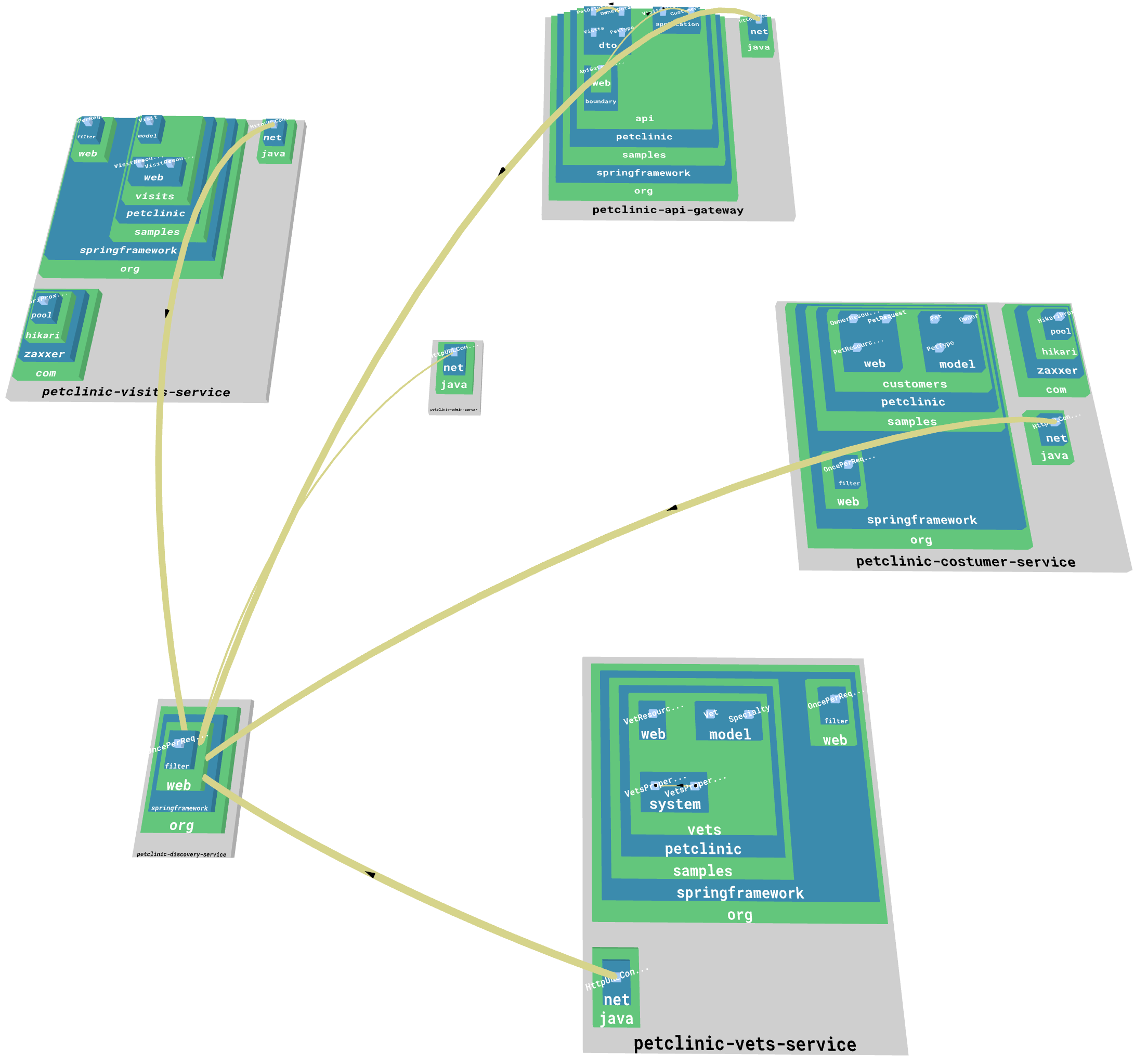}
	\caption{A visualization of the distributed PetClinic in the web interface of ExplorViz.
    Six applications are visualized with a grey foundation and hierarchically stacked packages in blue and green.
    Packages can contain classes. Communication between classes is visualized with yellow arcs.}
	\label{fig:explorviz}
\end{figure}

ExplorViz is our web-based software visualization tool which employs the city metaphor~\cite{fittkau2017,hasselbring2020,Wettel2007}.

The visualization focuses on the visualization of package structures, classes, and the live visualization of execution traces.
Dynamic program analysis is employed to collect runtime data, e.g., with NovaTec’s Java agent inspectIT Ocelot\footnote{\url{https://www.inspectit.rocks}} or Kieker~\cite{kieker2012,kieker2020,kieker2024,yang2025kieker}.
The traces are exported to a Collector component and need to adhere to the OpenTelemetry\footnote{\url{https://opentelemetry.io}} standard.
This enables ExplorViz to process traces from various sources, as long as the traces are compliant with OpenTelemetry's standard.
As long as method calls that make up a trace also carry the information about their class of origin, this suffices for ExplorViz to reverse engineer the structure of packages and classes.

In addition, static program analysis can be employed to collect data on classes for different software versions.
Thus, it can be tracked which classes have been added or removed throughout the software's evolution.

The backend of ExplorViz consists mostly of services that employ the Quarkus\footnote{\url{https://quarkus.io/}} framework with Java or Kotlin.
An exception is the collaboration service, which is written in Node.js.\footnote{\url{https://nodejs.org/en}}
The collaboration service supports multi-user collaboration for desktop computers, as well as for augmented and virtual reality devices~\cite{KrauseGlau2022}.
Messages to synchronize the visualization between multiple users are exchanged via WebSocket connections.
Messages between Quarkus services are exchanged via Apache Kafka\footnote{\url{https://kafka.apache.org/}}.

The web interface of ExplorViz is written in JavaScript, while three.js\footnote{\url{https://github.com/mrdoob/three.js}} is used as a library to render the 3D scene.
Figure \ref{fig:explorviz} illustrates the visualization in the frontend of ExplorViz of a distributed version of the PetClinic.\footnote{\url{https://github.com/spring-petclinic/spring-petclinic-microservices}}
Applications are visualized as cities with a gray foundation.
Packages are represented by green and blue boxes that represent districts in the city metaphor.
Classes are represented as blue buildings inside districts.
Text labels are placed on the visual entities with their corresponding name.

The yellow arcs between classes represent the accumulated method calls between the objects of the given classes.
Black arrows on the arcs indicate the direction of communication.
For applications, packages, classes, and communication, more detailed information is displayed in a popover when hovered with a mouse.

\section{Semantic Zoom}\label{sec:semantic-zoom}
\begin{figure*}
  \centering
	\includegraphics[width=\textwidth]{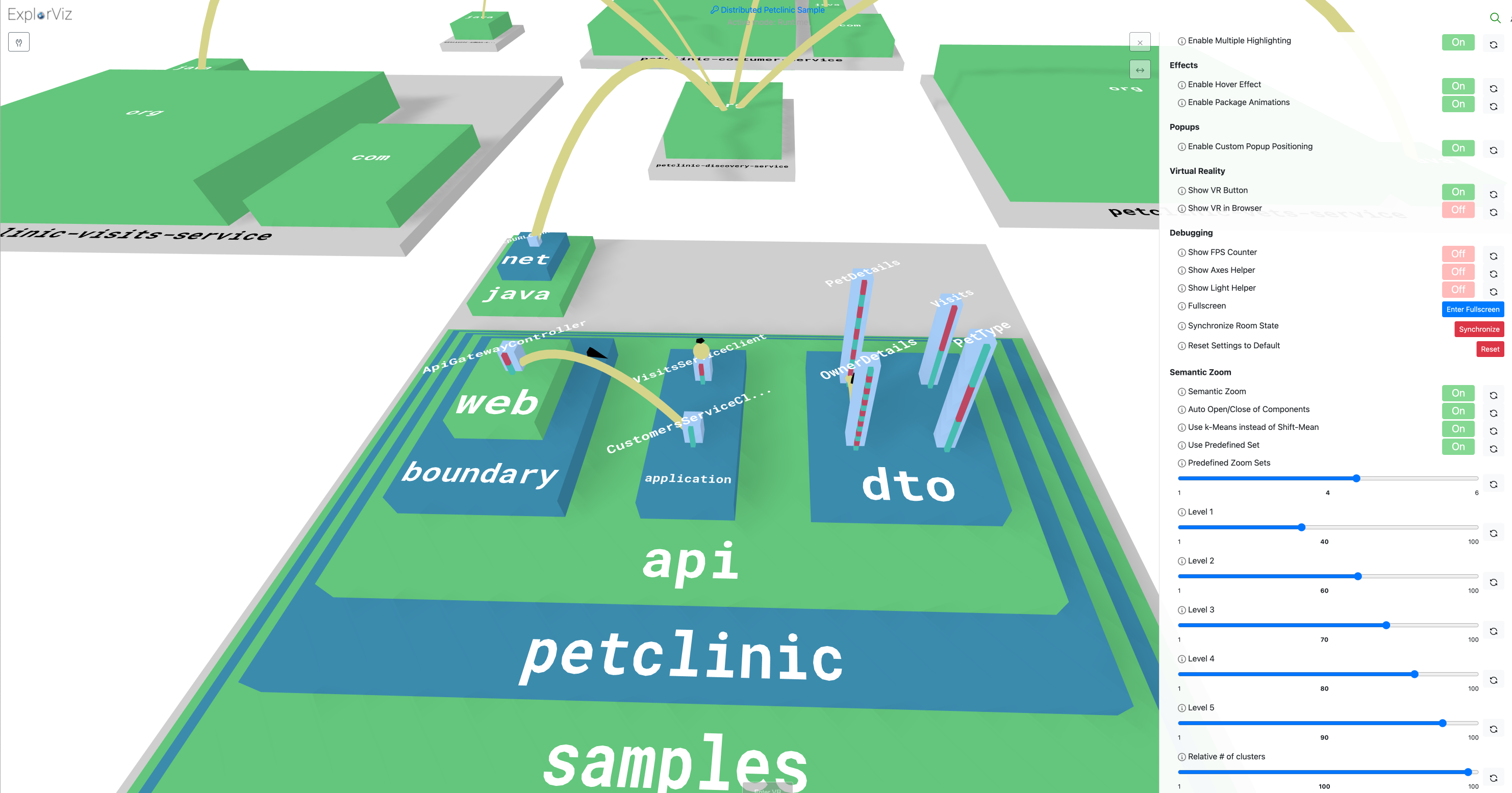}
	\caption{Visualization in ExplorViz with enabled semantic zoom feature. Among other changes, packages close to the camera are opened, classes are augmented with colored indicators for methods, and arrows on communication lines indicate the direction of the underlying requests. An opened settings component to the right allows to customize the semantic zoom. }
	\label{fig:semantic-zoom}
\end{figure*}

\textit{Definition and Categorization}
Semantic zoom has multiple definitions:
It is mostly compared with the geometric zoom or physical zoom, which only scales the objects, but does not present further information.
Semantic zoom is described to change the structure of objects that are to be displayed.
This may include changes in shape or appearance in any way to display other data on different spatial scales~\cite{semanticzoomwiki}.
Many related works that focus on semantic zoom do only consider 2D representation of data ~\cite{Wiens2017}, ~\cite{4015436}. 
Compared to semantic zoom, the term level of detail (LoD) refers to the mechanism of displaying visual objects with different numbers of polygons.
This is a common technique in complex 3D environments, such as in video games, to optimize performance.
There are three major LoD variants. 

First, there is discrete LoD.
The application receives multiple preprocessed versions of an object with a varying number of polygons. Subsequently, the software determines the optimal choice. As the angle of the object is not known in the preprocessing step, it is not possible to implement any optimizations in that regard.

The next significant LoD variant is the continuous LoD variant. 
The data structure for an object is streamed at runtime and provides enhanced granularity.
Refinements could also be streamed via a slow internet connection.

Lastly, there is view-dependent LoD. 
It is a combination of continuous LoD and viewing angle.
It is possible for larger objects to have a greater degree of granularity of polygons that are in closer proximity to the viewer, while simultaneously reducing the number of polygons in the far distance from the viewer.
However, despite this variation, the object remains coherent ~\cite{luebke2003level}.
 
In general, LoD focuses on the reduction of meshes and polygons for an object.
Even though the goals of semantic zoom and LoD differ, the categorization in discrete, continuous, view-dependent can be made for both.

\textit{Concept}
We intend to employ semantic zooms to be able to display more data when zoomed in and adapt or hide visual elements when zooming out.
Particularly, we add the following visual changes to ExplorViz:
\begin{enumerate}
    \item The height of classes can change. The metric of how many instances have been created of that class is taken into account.
    \item Meshes representing methods of a class are attached in a stack to the class mesh and adopt their height. The height of an individual method mesh represents the lines of code (LoC) in relation to the other methods of that class.
    \item Method meshes can be hidden.
    \item The size of class labels can change.
    \item With increasing size, labels will be shortened to avoid overlaps with other labels.
    \item The thickness of communication can be changed.
    \item The curvature of the communication can be adapted to make them more visible when zoomed out.
    \item Communication and their indicators for the direction of the communication can be hidden. This takes the number of requests into account that the communication visualizes.
    \item Packages can be closed automatically, thereby hiding all inner classes and subpackages. Labels of closed packages are centered and communication is aggregated for closed packages.
\end{enumerate}

These changes are designed to improve the overview when zoomed out.
In Figure \ref{fig:semantic-zoom}, a screenshot illustrates the distance-dependent visualization introduced by our semantic zoom approach.
The semantic zoom is an extension that aligns itself well with previously implemented filter options and popovers that show additional information when hovering over packages, classes, or communication links.
In combination, this follows the paradigm of ``overview first, zoom and filter, then details-on-demand,'' also known as Shneiderman's mantra~\cite{Shneiderman2003}.

\textit{Implementation}
As the appearance of objects should change with their distance to the camera, a change in the camera's position is the trigger for the (re-)computation.
However, the naive approach would require to compute the distance between the camera and almost all objects in the 3D scene.
Especially for large software landscapes, this is computationally demanding.
Therefore, we cluster all relevant 3D objects in advance and only compute the distance to the centroids of the clusters.

It can be configured whether clustering is performed with the k-Means~\cite{kmeans} or Mean Shift~\cite{meanshift} clustering algorithm.
Both of them are centroid-based clustering algorithms that automatically provide us with center points.
The number of clusters is a trade-off between performance and the spatial granularity of the semantic zoom, whereas the given landscape and performance of the user's hardware are important.
Therefore, users may configure the number of computed clusters in the web interface.

Every object that is able to change its appearance implements an interface for semantic zoom.
When the distance changes, objects are triggered and then each object can decide which appearance it takes (most of the time the appearance does no change at all).
Therefore, our implementation falls in the same category as discrete LoD~\cite{luebke2003level}.
As the distances at which changes in the appearance should occur may be influenced by different factors, such as the size of the monitor or user preference, a user can change which level of appearance is triggered at which distance.

\section{Mini-Map}\label{sec:minimap}
\begin{figure*}
  \centering
	\includegraphics[width=\textwidth]{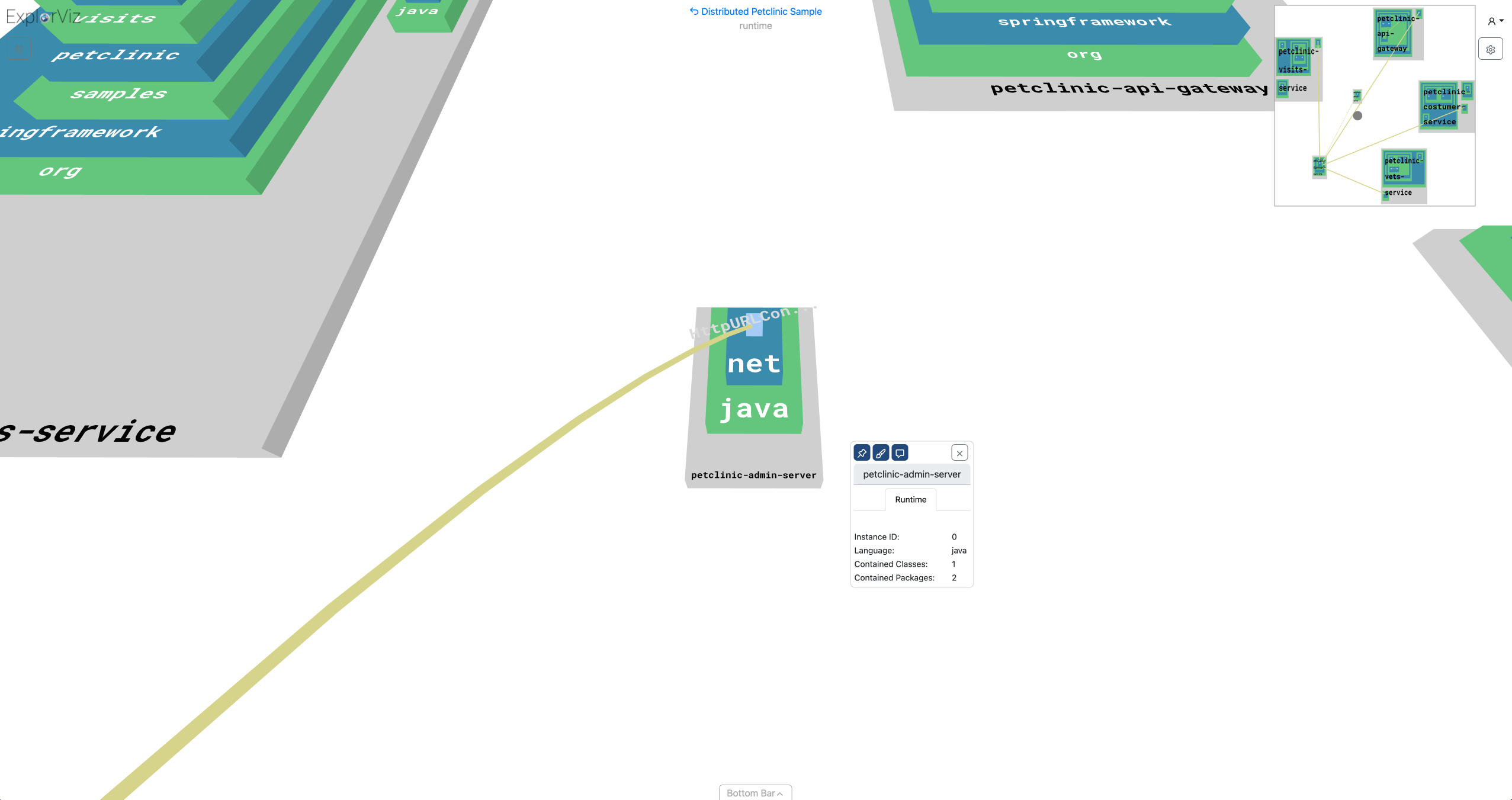}
	\caption{View of the distributed PetClinic, with the camera zoomed in on a single application. The mini-map in the top right corner provides a less detailed top-view of the entire software landscape.}
	\label{fig:minimap}
\end{figure*}

Mini-maps are commonly used in video games to aid navigating the world.
We adapted this concept and developed an implementation in our tool ExplorViz.
Figure \ref{fig:minimap} showcases our implemented approach with the distributed PetClinic.
The visual properties of the mini-map are mostly chosen in accordance with the results of Zagata and Medyńska-Gulij~\cite{mini-map-design}.
The shape is rectangular which aligns the mini-map with the mostly rectangular and axes-aligned visual elements of ExplorViz.
We position the mini-map in the top right corner, since a placement at the bottom of the page would interfere with our timeline for selecting a timestamp.
A static orientation instead of a player-centric approach is used since the visual elements have a static orientation and it allows us to display textual hints which do not need to be rotated.

For centering, we are mostly world-centered to keep the mini-map mostly static.
In addition, we ensure that always some part of the software landscape is visible in the mini-map to prevent a ``whiteout.''
The employed projection is orthographic, such that our rectangular elements are not affected by perspective skews.
The base layer is usually artificial in mini-maps.
Since ExplorViz already uses a simplified view and no complex 3D meshes, we stick with the original elements but hide or adapt labels to fit the mini-map.
Our mini-map takes up around 4\% of the screen, which is larger than in most video games.
However, this allows us to display more detailed information on the mini-map and, in our experience, does usually not occlude relevant elements in ExplorViz.

We do not employ specific navigational elements, as the mini-map is always oriented in the same direction.
As a visual clue, we indicate the user's position with a gray circle.
We offer two options for the position of this circular indicator, since we use the \textit{OrbitControls} of three.js, i.e. the user rotates its camera around a point called target.
Thus, we allow to display either the camera's or target's position, so a user can determine their own position or the position of what they are looking at in the mini-map.

\begin{figure*}
  \centering
	\includegraphics[width=\textwidth]{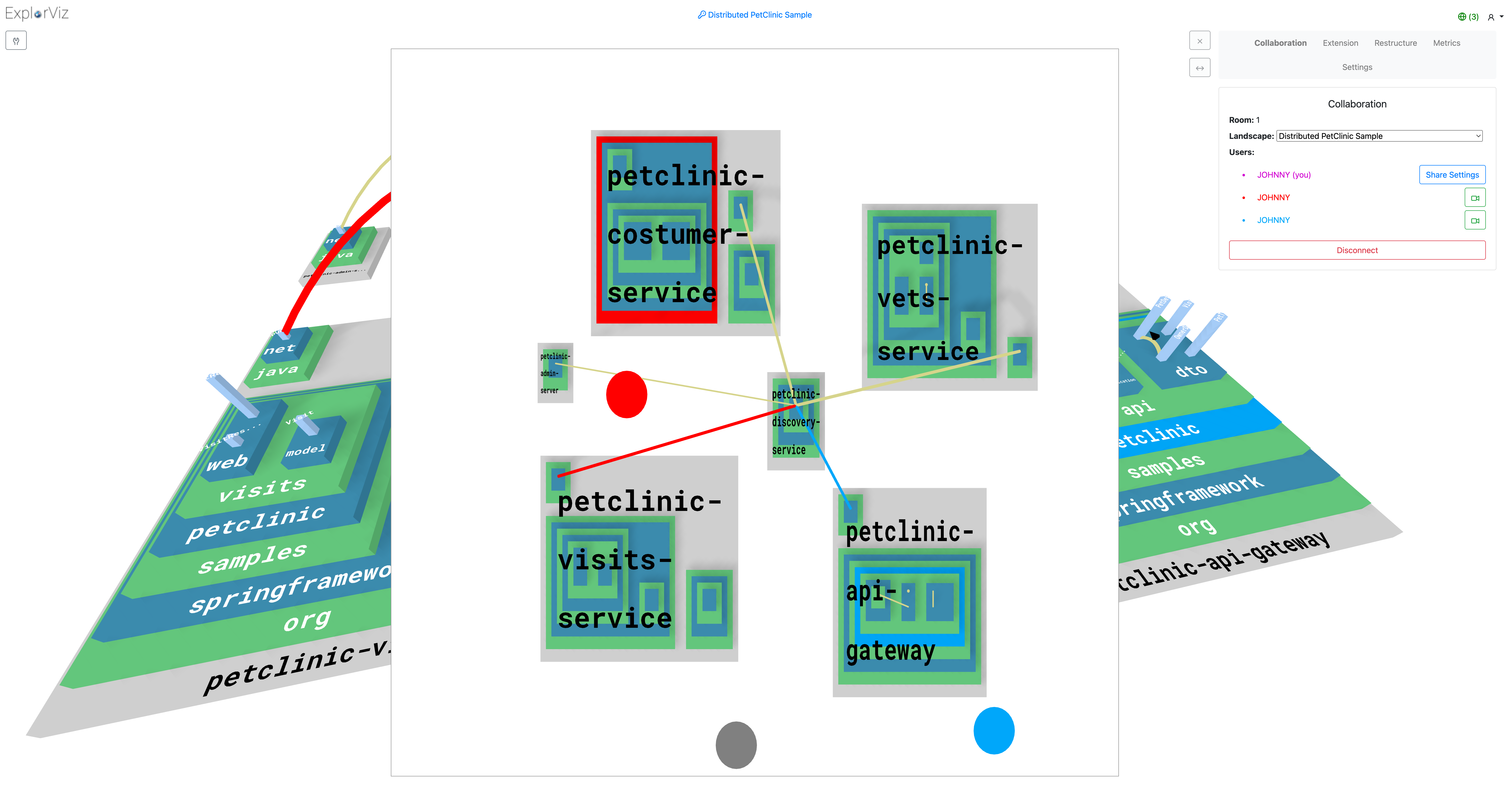}
	\caption{The mini-map can be temporarily enlarged. This provides an orthographic view of the software landscape. Colorings show what was highlighted by the connected users that are displayed as colored circles. A web component with an overview of connected users is opened to the right.}
	\label{fig:maximap}
\end{figure*}

In a settings component, a configurable zoom enables the user to determine how much of their surroundings should be visualized in the mini-map.

Aside from the orthographic view and adapted labels, the visualization mostly resembles the original top-view of ExplorViz.
As an addition, the own position and the position of other color-coded users is indicated with circles on the mini-map.
These visual markers can be clicked anytime to teleport to that user's location, that is, the camera is placed at the same position and in the same orientation as the other user's camera.
Combined with the ability to colorize (highlight) visual elements, we assume that this may help the crucial aspect of collaboration in our software visualization~\cite{Roehm2012,Maalej2014,KrauseGlau2022}.

A mouse click on the mini-map enlarges it to fill most of the screen (see Figure \ref{fig:maximap}).
This view is suitable to observe the state of the landscape and movements of other users.
The enlarged mini-map leaves enough space, such that other frontend components can be opened without occlusion on either side.
In the given example, an overview of the virtual room is placed to the right of the mini-map.
Clicking on the camera next to another user's name allows to spectate that user, that is, the camera position and orientation and continuously updated accordingly.


Regarding the implementation, a separate orthographic camera is placed above the landscape.
The additional view is rendered on a separate canvas that is scissored\footnote{\url{https://developer.mozilla.org/en-US/docs/Web/API/WebGLRenderingContext/scissor}} into the original canvas.
Elements that should be hidden for the mini-map or are specifically added are placed into separate visual layers.
Therefore, our implementation is robust to changes in the underlying visualization and only requires to set an appropriate layer for newly added elements that should not be part of the mini-map visualization.

\section{Evaluation}\label{sec:evaluation}
\begin{figure}[htbp]
  \centering
  \begin{subfigure}[b]{0.24\textwidth}
    \centering
    \includegraphics[width=\textwidth]{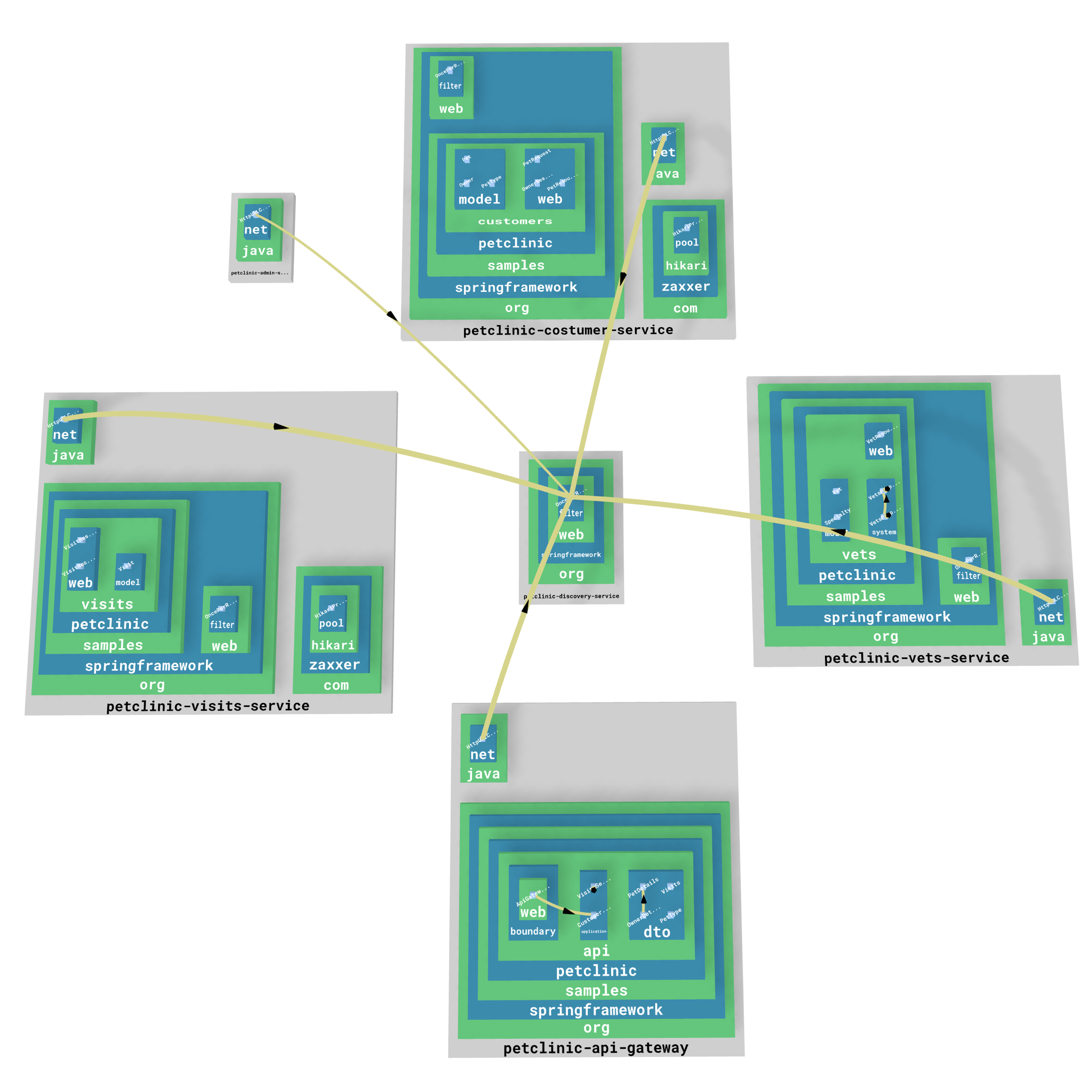}
    \caption{Distributed PetClinic}
  \end{subfigure}
  \hfill
  \begin{subfigure}[b]{0.24\textwidth}
    \centering
    \includegraphics[width=\textwidth]{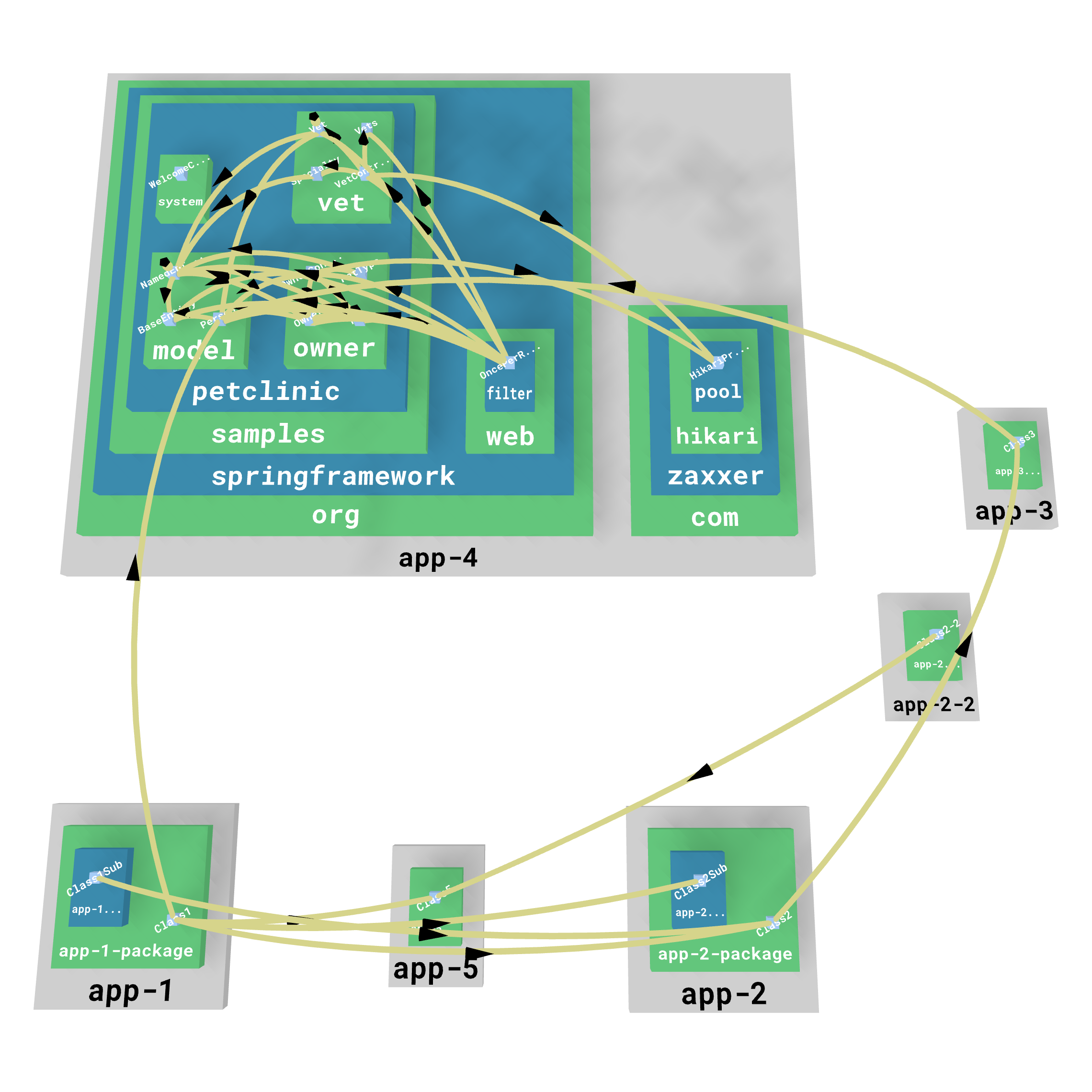}
    \caption{Artificial Landscape}
  \end{subfigure}

  \vspace{1em}

  \begin{subfigure}[b]{0.24\textwidth}
    \centering
    \includegraphics[width=\textwidth]{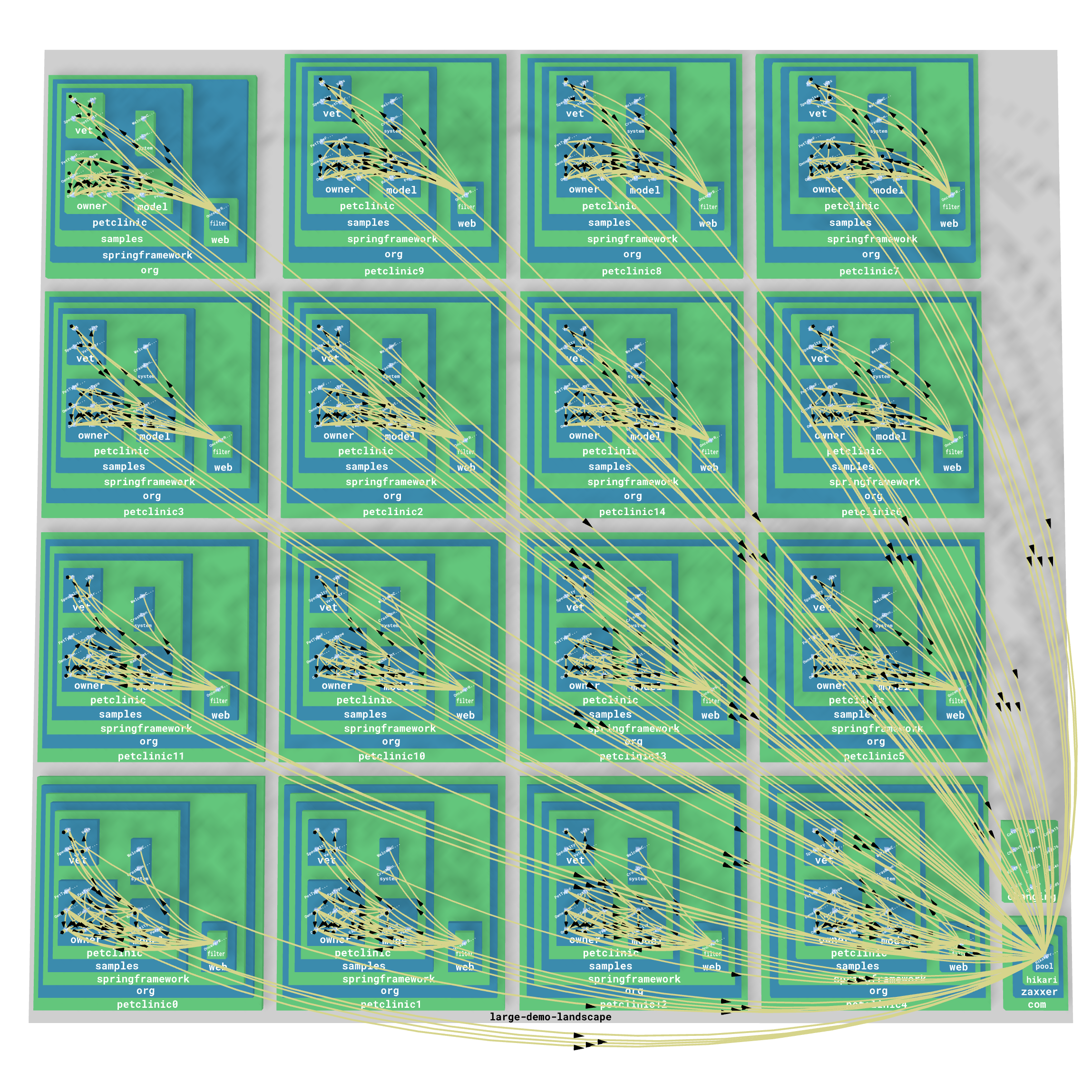}
    \caption{Large Landscape}
  \end{subfigure}
  \hfill
  \begin{subfigure}[b]{0.24\textwidth}
    \centering
    \includegraphics[width=\textwidth]{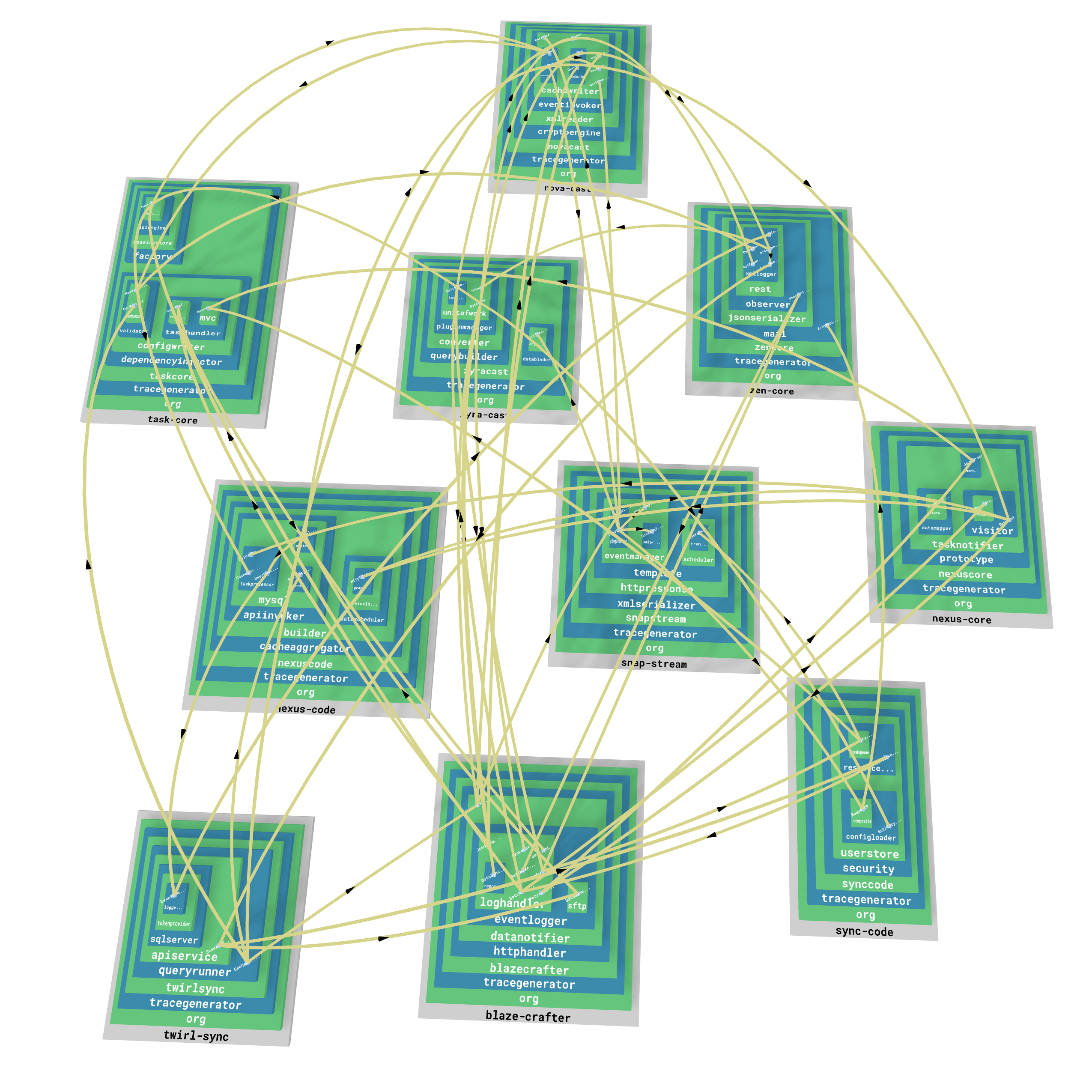}
    \caption{TraceGen XL}
  \end{subfigure}

  \caption{Overview of the software landscapes and their names that were used in our evaluation.}
  \label{fig:software-landscapes}
\end{figure}

The semantic zoom and mini-map feature have been evaluated in isolation and about three months apart from one another.
Different software landscapes were used for the evaluation.
An overview of the landscapes is given in Figure \ref{fig:software-landscapes}.
Figure \ref{fig:software-landscapes} (a) is a snapshot of the distributed PetClinic, Figure \ref{fig:software-landscapes} (b) is the Spring PetClinic that has been extended with artificial applications, Figure \ref{fig:software-landscapes} (c) shows a PetClinic with duplicated packages, Figure \ref{fig:software-landscapes} (d) is a completely synthetic landscape that has been generated with our trace generation tool.\footnote{\url{https://github.com/ExplorViz/trace-generator}}
We make all results of our surveys, the employed software artifacts, and an accompanying video publicly available~\cite{hansen_2025_15491590}.

In the upcoming sections, we first present the semantic zoom study and then go on to the mini-map evaluation.

\subsection{Semantic Zoom Evaluation}\label{sec:semantic-zoom-evaluation}
For the evaluation of our implemented semantic zoom feature, we are motivated by two research questions.
\begin{itemize}
   \item \textbf{RQ1} \textit{Does the semantic zoom feature increase productivity?}
   \item \textbf{RQ2} \textit{Does the semantic zoom feature improve the user experience?}
\end{itemize}

The published dataset of the study also contains results on a feature that allows users to explore class data in an immersive way~\cite{hansen_2025_15491590}.
However, we do not consider that feature in this paper.

\textbf{Setup}
All participants were invited to a computer laboratory at Kiel University.
They used a Windows computer with an Intel Core i5-6500 CPU, a NVIDIA GeForce GTX 1070 GPU, and a 24-inch monitor with full-HD resolution.
A second monitor was used to display the LimeSurvey tool.
Google Chrome was employed as a browser to fulfill the tasks in ExplorViz.

\textbf{Participants}
16 people, who were mostly computer science students, took part in the evaluation.
Eight participants have used ExplorViz before and eight participants had no prior experience with ExplorViz.
Among the participants with prior experience in ExplorViz may be participants who also took part in the evaluation of the mini-map.
In accordance with Wiens et al., each participant was assigned to a group, referred to as group A and group B~\cite{Wiens2017}.
Group A was made up of nine participants, whereas seven participants were assigned to group B.
There was an additional participant in group B, for whom the survey tool timed out and thus no complete data set could be collected.
For the sake of completeness, this participant's partial data is included in the accompanying dataset but is not considered in this paper.

\textbf{Methodology}
The study participants were asked to complete a questionnaire and work on the presented tasks.
First, the participants answered questions on their background, e.g. experience with software development and ExplorViz.
Afterwards, each participant received an introduction to ExplorViz and could explore the semantic zoom features.
In the following main part of the study, the participants were asked to fulfill four tasks in ExplorViz.
\begin{itemize}
    \item Task 1 asked to find a class in the ``Distributed PetClinic'' software landscape and name the underlying application.
    \item Task 2 asked to name all classes of a subpackage in our ``Artificial Landscape.''
    \item Task 3 asked which classes might be affected by a change in a class called "Vet" in the "Artificial Landscape".
    \item Task 4 asked to determine the direction of a communication in the ``Tracegen - XL'' software landscape.
\end{itemize}
Participants of group A solved Tasks 1 and 2 with enabled semantic zoom feature and Tasks 3 and 4 without semantic zoom enabled.
For participants of group B, the use of semantic zoom was reversed, i.e., semantic zoom was only enabled for Tasks 3 and 4.
Different software landscapes were employed such that the participants could make use of the semantic zoom features in different scenarios with previously unknown applications.
The tasks are designed such that participants with different background can solve them, but all require the exploration of the software landscape, thereby triggering features of our implemented approach.
The time needed to fulfill the tasks was tracked and entered into the questionnaires.

In the end, the participants rated their experience and were also asked to provide textual feedback.

\textbf{Results}
As the tasks were designed to be solvable and foster interaction with the visualization, almost all given answers were correct.
Only the class names that appear in the answers for Task 3 vary. 

\begin{figure}[htbp]
	\includegraphics[width=\textwidth/2]{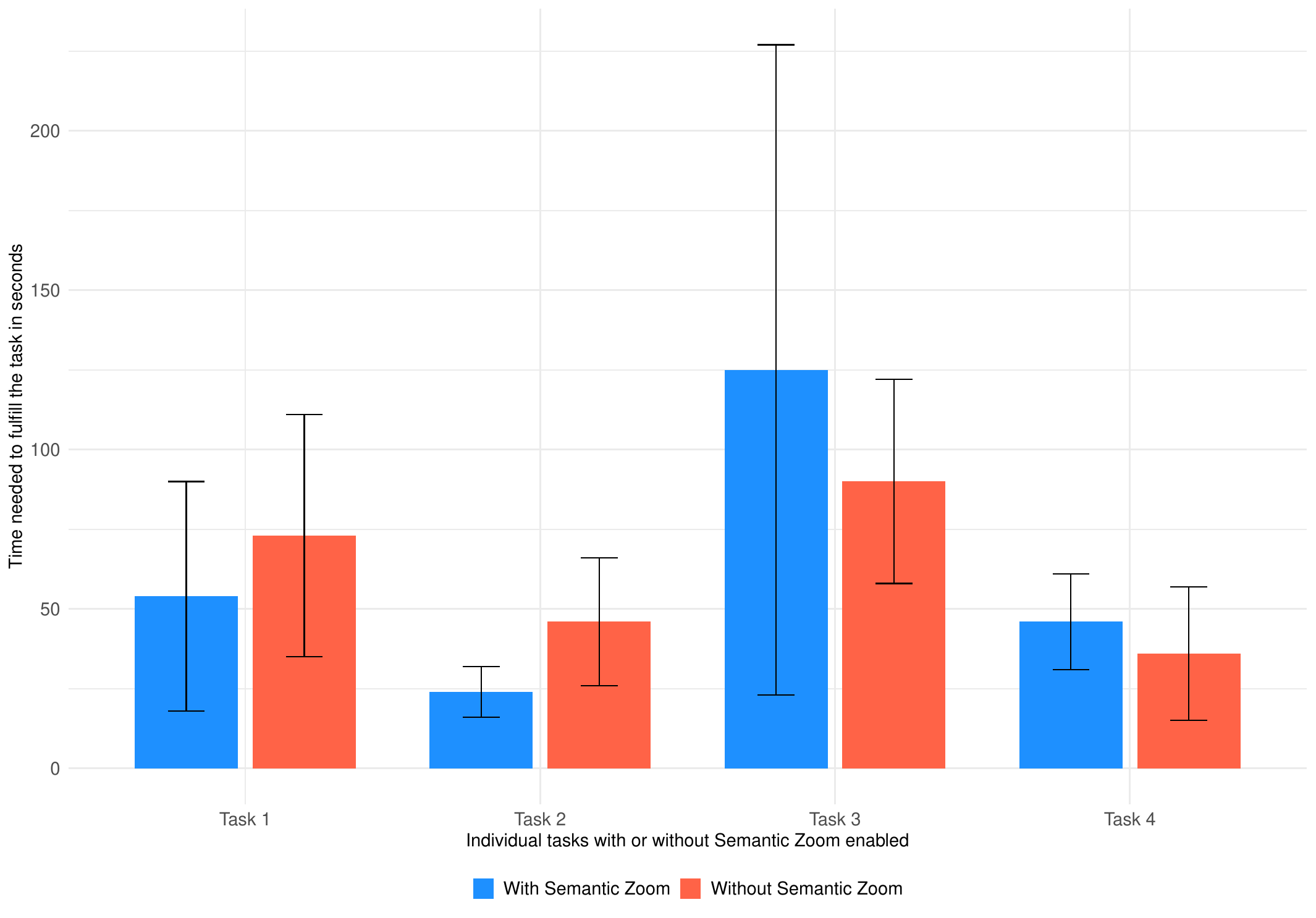}
	\caption{Mean times in seconds and their standard deviations for each task with and without semantic zoom enabled.}
	\label{fig:semantic-zoom-times}
\end{figure}

Figure \ref{fig:semantic-zoom-times} illustrates the recorded times for the 4 tasks with regard to the semantic zoom feature.
It can be seen that Task 3, which asked about implications for a refactoring task, took the longest overall.
The comparison of times between participants who used semantic zoom and those who did not, is inconclusive.
Participants using semantic zoom completed Tasks 1 and 2 faster, while the overall times are higher for Tasks 3 and 4 with sematic zoom.
Notably, the blue bar for Task 3 contains an outlier who took 364 seconds to complete the task, whereas the other 16 participants finished in between 37 and 145 seconds.
The wide range of results is indicated by the standard deviation, visualized as black lines in Figure \ref{fig:semantic-zoom-times}.

\begin{figure}[htbp]
	\includegraphics[width=\textwidth/2]{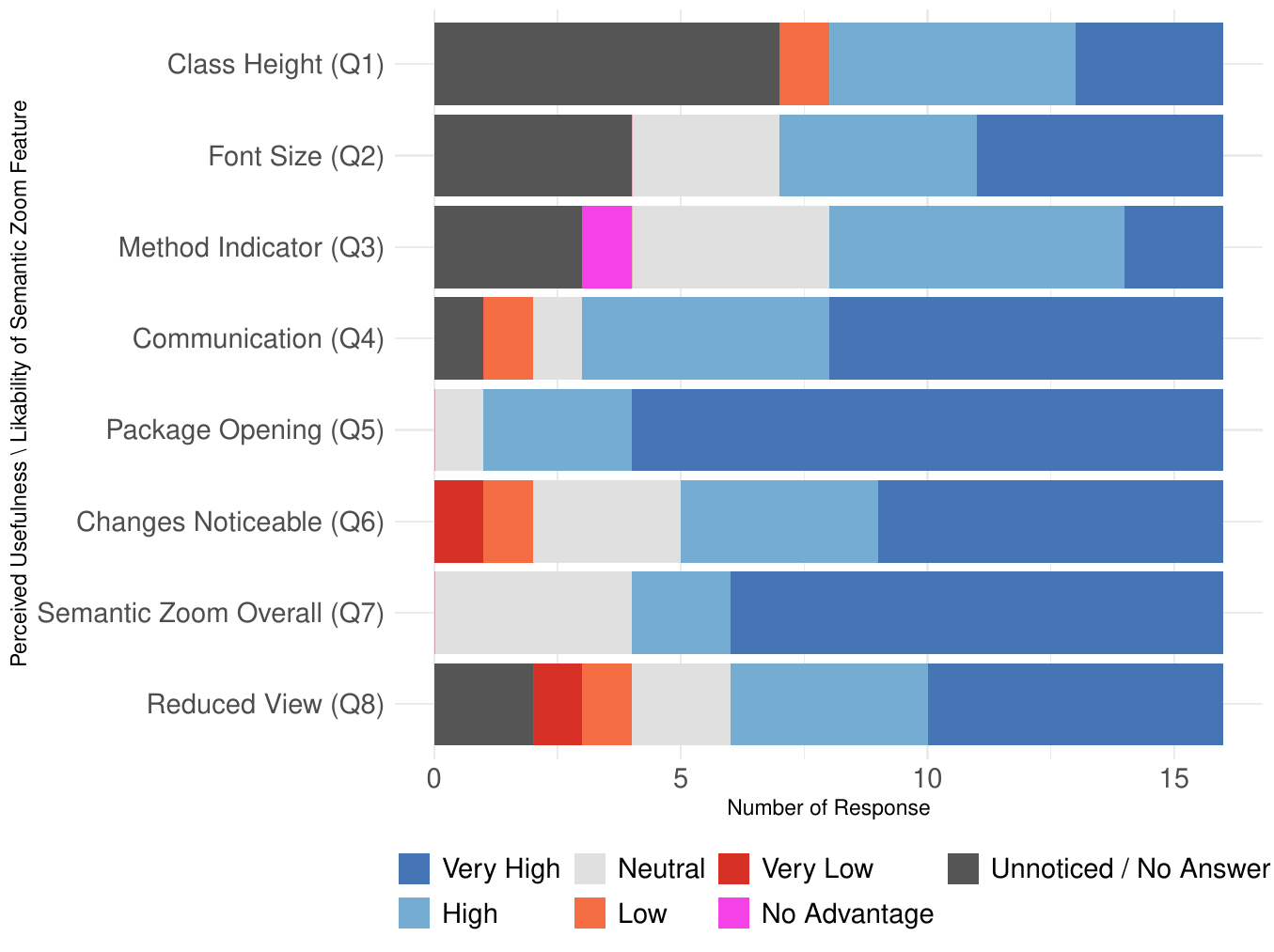}
	\caption{Overview of results regarding perceived usefulness and likability of different features, including the semantic zoom approach as a whole.}
	\label{fig:semantic-zoom-usefulness}
\end{figure}

Figure \ref{fig:semantic-zoom-usefulness} aggregates the results regarding usefulness and likability of the added semantic zoom features.

The usefulness of changing the class height (Q1) was perceived as high or very high by eight people, while seven people either did not notice the change or could not rate its usability.
Textual feedback suggested to change the color of classes that are in focus and increase the length of the displayed class label.
The automatic change of the font size of labels (Q2) was perceived as useful by nine participants.
One participant suggested increasing the font size for closed packages that do not display inner classes and packages.
The added method meshes (Q3) were perceived as useful by 8 participants, while one participant did not see it as an advantage.
Regarding Communication, 13 out of 16 participants liked that communication shrinked when zooming in.
One participant added as a feedback that the arrows to indicate the direction of the communication should always be visible.
All but one participant liked the automatic opening and closing of packages (Q5).
Textual feedback pointed out that this feature changes the visualization a lot and that it might be necessary to change semantic zoom settings to avoid that packages are opened or closed unexpectedly.
Question 6 asked if visual changes when zooming were noticed and expected.
This was the case for eleven participants, while two participants disagree.
Semantic zoom was rated as useful for daily life usage (Q7) by twelve participants, four had a neutral opinion.
The last question of Figure \ref{fig:semantic-zoom-usefulness} asked participants to rate how much they liked the view with closed packages compared to the regular view of ExplorViz.
The majority liked the reduced view, while two disliked it.

\begin{figure}[htbp]
	\includegraphics[width=\textwidth/2]{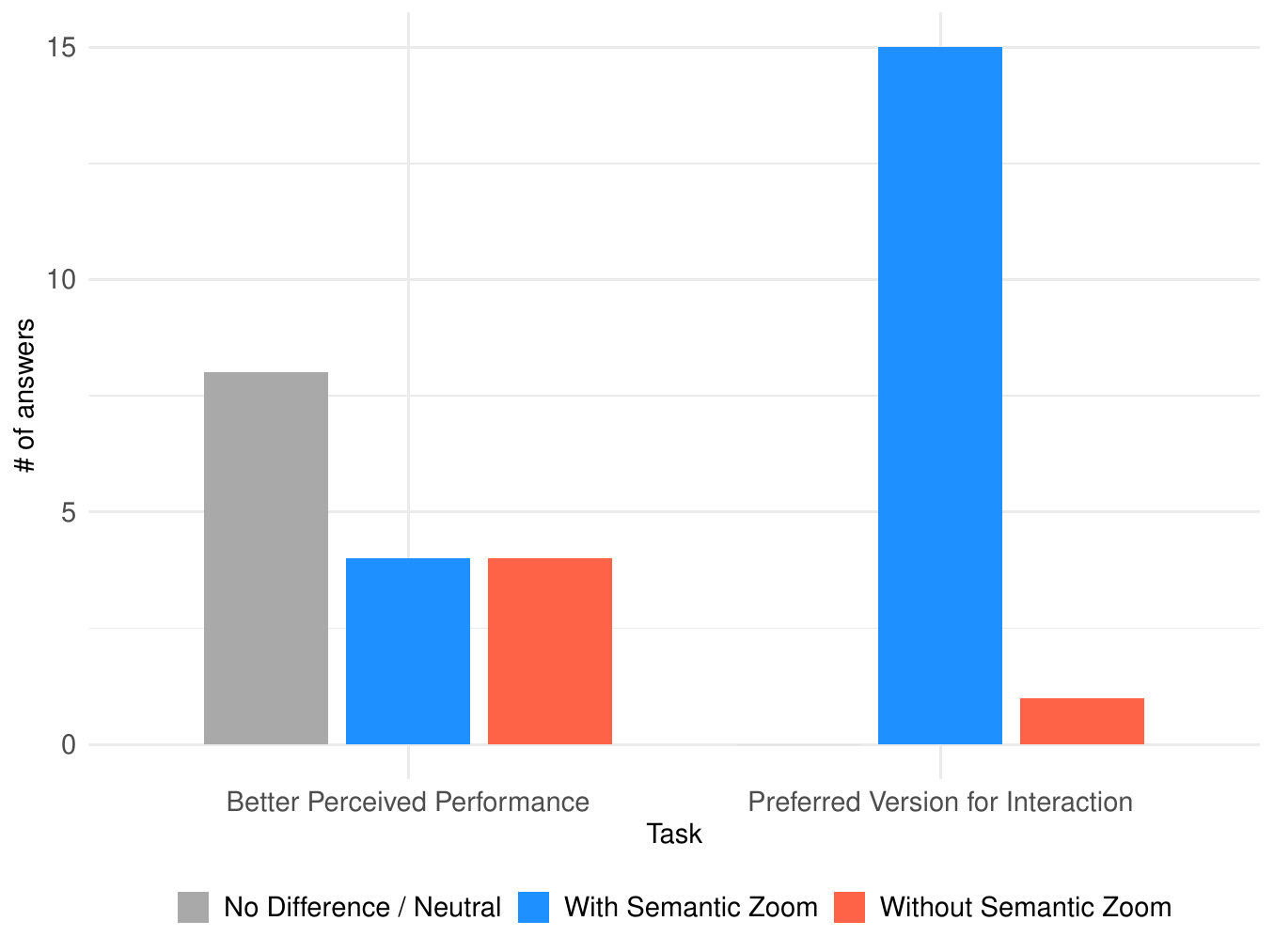}
	\caption{Semantic zoom feature compared to regular version of ExplorViz in terms of perceived performance and preferred version for interaction in the 3D visualization.}
	\label{fig:semantic-zoom-overall}
\end{figure}

Figure \ref{fig:semantic-zoom-overall} presents the results regarding the question whether the participants of our study liked the version with or without semantic zoom more in terms of performance and for interaction with the 3D visualization.
Regarding performance, most participants perceived no difference, and both for and against semantic zoom received the same number of answers.
Overall, 15 participants stated that they prefer the version with semantic zoom, while 1 participant liked the regular version more.
No participant chose to answer neutrally for this question.

\textbf{Discussion}
Regarding our first research question, we could not show that our implementation increases productivity in software visualization tasks.
The results in Figure \ref{fig:semantic-zoom-times} exhibit a large standard deviation such that no substantial time difference can be observed in favor or against our approach for semantic zoom.
We assume that other factors, including the varying knowledge of participants regarding software development and ExplorViz, in combination with the limited sample size overshadow any effect on productivity that our approach might have.

The overall positive ratings of usefulness and usability show us that semantic zoom can have a positive impact on the user experience in 3D software cities.
However, the individual ratings presented in Figure \ref{fig:semantic-zoom-usefulness} and related textual feedback shows us that it is difficult to notably change the appearance of a visual object without being distracting to the user.
We presume that a mature animation system and continuous changes to visual elements while preserving the user's mental map are crucial.

The results for perceived performance are in line with a preliminary performance test by us, which indicates that the version with semantic zoom has a slightly higher overall frame rate but exhibits drops in the frames per second more often than the regular version, thus, making it unclear which of the two implementations is better. 

\textbf{Threats to Validity}
The participants of the study were mostly computer science students and thus are not representatives of the group of professional software developers.
There is an overlap of participants in this study and the previous study on mini-maps.
However, we expect that the learning effect and its influence on the study results is negligible due to the considerable time gap between the two studies.
The relatively small sample of 16 people limits the significance of our results.
Even though the participants did not receive any form of compensation, many participants knew the conductor of the conductor of the study personally.
This might induce a bias in favor of our approach.

The employed software landscapes and tasks were designed to foster interaction with the visualization and be solvable without prior knowledge about ExplorViz.
Thus, this study did not consider complex and arguably more realistic tasks that go beyond the visual exploration of the employed software landscapes.

\subsection{Mini-Map Evaluation}
Just as the semantic zoom feature, ExplorViz extended by the mini-map implementation was evaluated in a user study.
The evaluation of the mini-map feature is motivated by the question whether mini-maps are a useful addition to 3D software cities.

\textbf{Setup}
The hardware setup is identical to the setup described in Section \ref{sec:semantic-zoom-evaluation}.

\textbf{Participants}
14 people with a background in computer science, either as a student or researcher, participated in the study.
Seven participants stated to have no prior experience with ExplorViz, while 5 participants had some experience and two participants ordinary or much experience.
Participants participated in the study in groups of two.
However, the participants each filled their own questionnaire independently of one another.

\textbf{Methodology}
At the beginning of the evaluation, participants were introduced to ExplorViz.
Following, the participants were asked to explore all the features of the mini-map in the ``Distributed PetClinic'' software landscape on their own.
This included the testing of different configurations in the settings for the mini-map.
Participants were always allowed to ask questions about ExplorViz and the implemented feature.
After this first part, participants were asked to rate the usability of the mini-map and its perceived usefulness for the given software landscape.
In a second part of the evaluation, two participants joined a collaborative session in ExplorViz.
For this part, they were again asked to use the implemented features of the mini-map, this time with a focus on collaboration.
Finally, the participants rated their experience regarding the collaborative use of the mini-map and were asked to give additional textual feedback. 

\textbf{Results}
The study was conducted in German.
Therefore, we translated the survey and the textual feedback for this section.

\begin{figure}[htbp]
	\includegraphics[width=\textwidth/2]{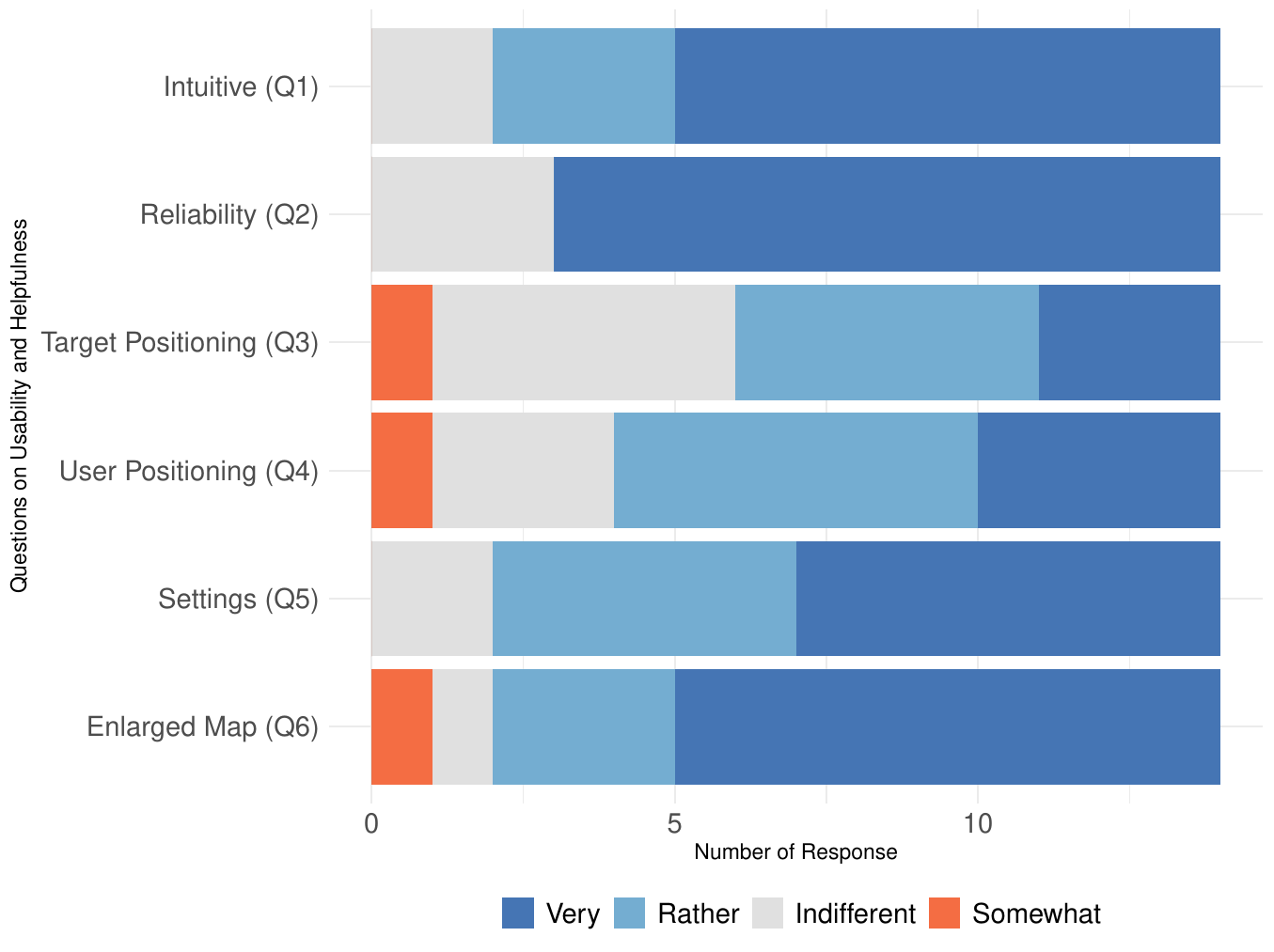}
	\caption{Overview of quantitative feedback regarding various features and aspects of the implemented mini-map.}
	\label{fig:mini-map-usablility}
\end{figure}

In Figure \ref{fig:mini-map-usablility}, the quantitative feedback that was collected after a participant explored the mini-map feature alone is presented.
When asked how intuitively the mini-map could be used (Q1), no participant perceived it as non-intuitive.
In the textual feedback, it is stated that the mini-map helps orientation as it is clear where the camera is currently positioned.
Asked about the reliability of the mini-map (Q2), most participants were very satisfied.

As ExplorViz uses \textit{OrbitControls}, the marker indicating the current position on the mini-map could be set to the user's position or to the position of the target that is orbited.
Comparing these options (Q3 and Q4), the given ratings are very similar.
One participant mentions that the positioning on the target makes no sense to him/her, but that it might be useful for other users.
When asked about how helpful the provided settings are (Q5), twelve of 14 participants rated it as rather or very helpful.
One participant stated that being able to control which kind of elements are displayed could be helpful for large projects.
The enlarged mini-map was perceived as rather or very helpful by twelve people, two people are indifferent or think it is rather not helpful.
The textual feedback mentions that the enlarged mini-map is not intuitive and lacks interactive features.
One participant suggested to display additional text whenever elements are hovered.

The remaining feedback for the first part includes suggestions to allow users to teleport to a given point by clicking on the mini-map and do not apply zoom to both the mini-map and enlarged mini-map.

\begin{figure}[htbp]
	\includegraphics[width=\textwidth/2]{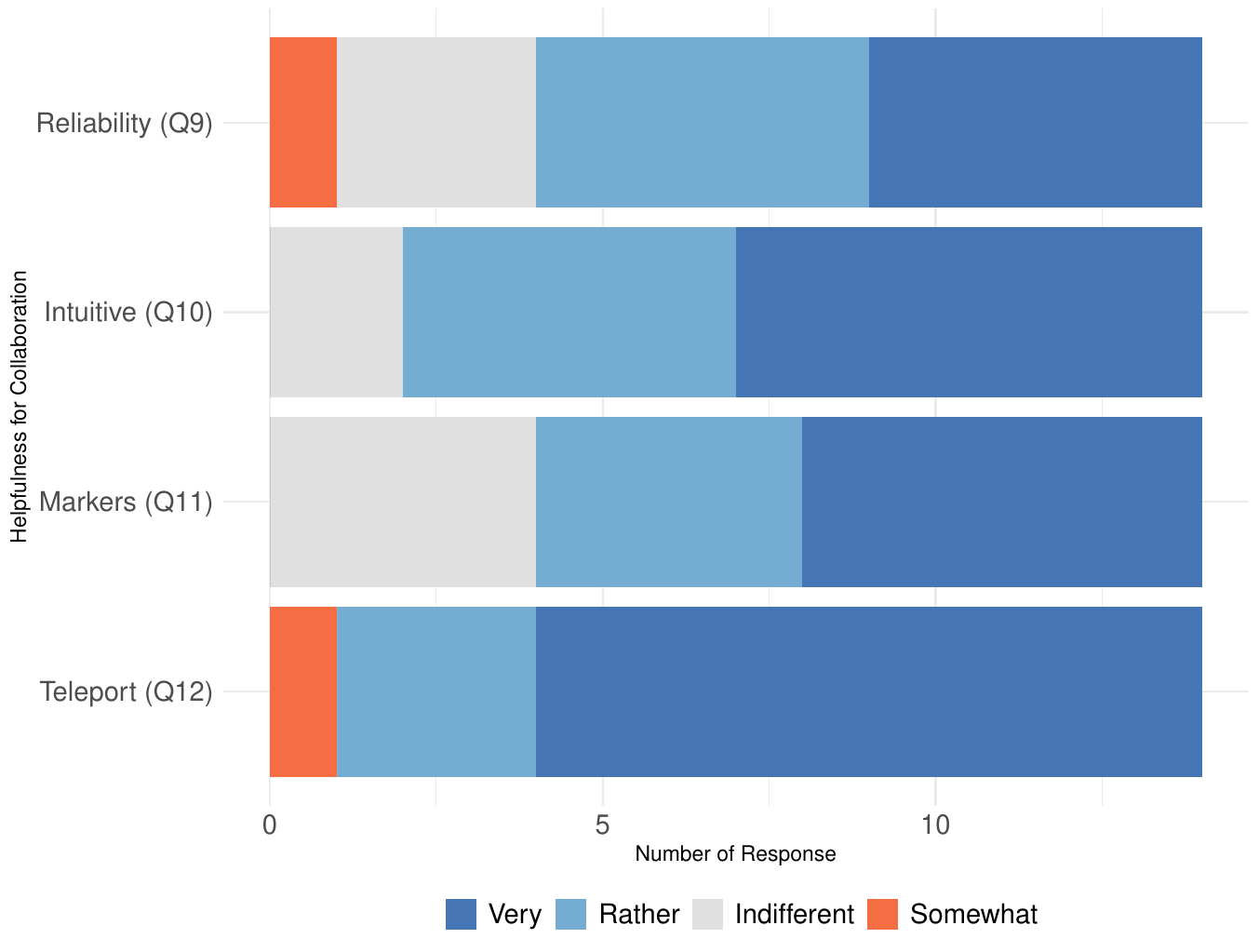}
	\caption{Quantitative user feedback on collaborative use and features of the mini-map.}
	\label{fig:mini-map-collaboration}
\end{figure}

In Figure \ref{fig:mini-map-collaboration}, the quantitative feedback regarding the collaborative use of the mini-map is illustrated.
The reliability for collaboration (Q9) is rated worse than for the single-user task (Q2).
The textual feedback makes it clear, that visualization of other user's markers did not work all the time and some participant needed to reload the webpage to make it work.
The intuitive use (Q10) of the mini-map is also rated slightly worse than before (Q1).
The textual feedback mostly criticizes that the markers for other users are not shown when they are outside the mini-map.
In addition, the marker for other user's always show the user's position whereas the own user's marker also could show the target's position.
Nonetheless, the markers for user positions (Q11) were rated as rather or very helpful by 10 participants.
The teleportation to other user's position by clicking on their marker on the mini-map (Q12) was perceived as rather or very helpful by 13 participants.

\begin{figure}[htbp]
	\includegraphics[width=\textwidth/2]{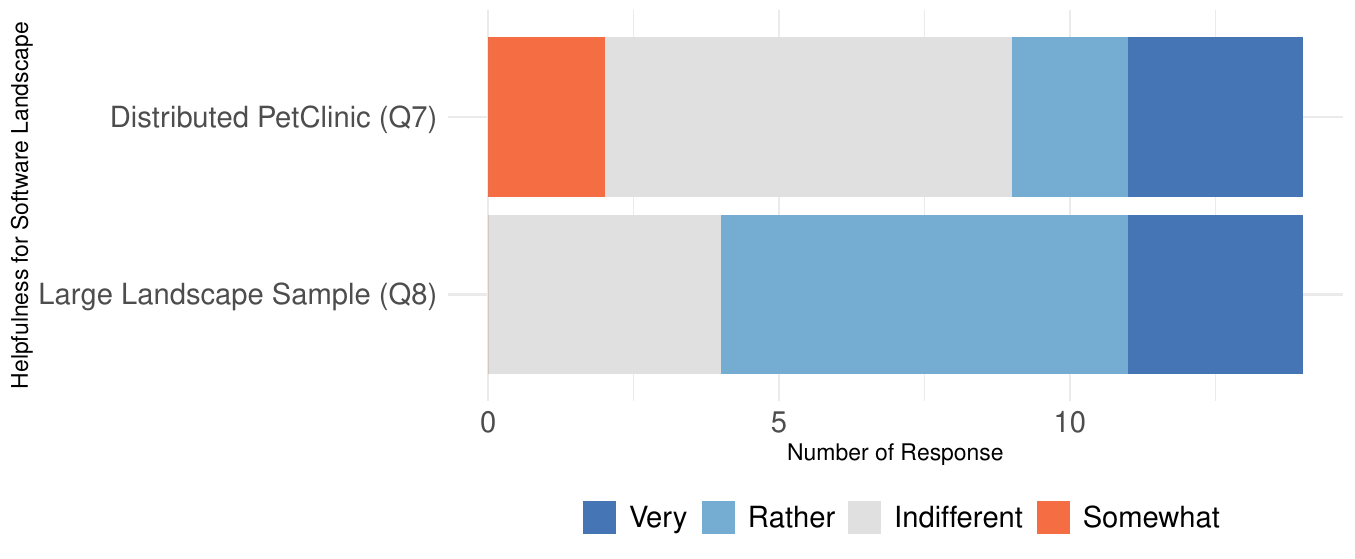}
	\caption{User feedback on the helpfulness two different software landscapes.}
	\label{fig:mini-map-landscapes}
\end{figure}

In Figure \ref{fig:mini-map-landscapes}, the results for perceived helpfulness of the mini-map for the two landscapes that were employed in the evaluation are illustrated.
It can be seen that the mini-map is perceived as more helpful in the ``Large Landscape Sample,'' which has more visual elements than the ``Distributed~PetClinic.''

\textbf{Discussion}
We presume that the common use of mini-maps in video games and adhering to established conventions regarding the design of the mini-map, facilitated the use of the mini-map for our study participants.

The indication that the mini-map feature is more helpful in larger, more complex software landscapes (see Fig. \ref{fig:mini-map-landscapes}) meets our expectations.
For software landscapes that only consist of a single and simple application, the benefits for navigating the visualization can be expected to be negligible.

Regarding the collaborative use of the mini-map, the feedback was mostly positive.
The study revealed that our implementation has some shortcomings and did not work reliably in all situations.
However, the participants gave mostly positive feedback and stated that the collaborative features may be useful for the collaborative exploration of large landscapes.

Overall, the participants mostly criticized shortcomings in the implementation of the mini-map, no participant stated that the mini-map is obstructive or not helpful at all.
Therefore, we conclude that mini-maps can be a useful addition to visualizations with 3D software cities.

\textbf{Threats to Validity}
Again, the participants of the study were mostly computer science students and thus are not representatives for the group of professional software developers.
The sample size of 14 people is not large enough to generalize our study results.
Many participants knew the conductor of the study which might induce a bias in favor of the mini-map approach.

\section{Conclusions and Future Work}\label{sec:conclusion}
We introduced approaches for semantic zoom and mini-maps for 3D software cities using the example of ExplorViz.
The semantic zoom changes the visual appearance of displayed elements and can reduce visual complexity depending on the camera's distance to the elements.
The mini-map offers an overview of the software landscape in the top right corner.

We evaluated our approaches in two studies.
The collected feedback from study participants is mostly positive.
This indicates that the concepts of semantic zoom and the mini-map, known from related research and video games, are also applicable to visualizations for 3D software cities.

There are many ways in which our approach could be further extended.
For semantic zoom, we plan to explore a more mature animation system and smooth transitions to change the appearance of more visual elements.
The mini-map could be extended with further visual indicators and offer more visual customization options.
Both semantic zoom and the mini-map could be valuable additions for software exploration collaborative virtual reality environments.
To gain further insight, an evaluation of a refined implementation with software professionals would be desirable.

\section*{Acknowledgments}
We want to thank the participants of the conducted studies for their time and valuable feedback.

\providecommand{\doi}[1]{DOI: \href{https://doi.org/#1}{#1}}
\bibliographystyle{myIEEEtran}
\bibliography{bibliography}

\end{document}